# Discrete superconducting phases in FeSe-derived superconductors

T. P. Ying[1,†], M. X. Wang[1,†], Z. Y. Zhao[1], Z. Z. Zhang[2], X. Y. Jia[1], Y. C. Li[1], B. Lei[3], Q. Li[1], Y. Yu[1],

E. J. Cheng[1], Z. H. An[1,4], Y. Zhang[1,4], W. Yang[5], X. H. Chen[3,4*] and S. Y. Li[1,4*]

[1] *State Key Laboratory of Surface Physics, Department of Physics, and Laboratory of Advanced Materials, Fudan University, Shanghai 200438, China*

[2] *Key Laboratory for Renewable Energy, Beijing Key Laboratory for New Energy Materials and Devices, Beijing National Laboratory for Condensed Matter Physics, Institute of Physics, Chinese Academy of Sciences, School of Physical Sciences, University of Chinese Academy of Sciences, Beijing 100190, China*

[3] *Hefei National Laboratory for Physical Sciences at Microscale and Department of Physics, and Key Laboratory of Strongly-coupled Quantum Matter Physics, Chinese Academy of Sciences, University of Science and Technology of China, Hefei, Anhui 230026, China*

[4] *Collaborative Innovation Center of Advanced Microstructures, Nanjing 210093, China*

[5] *Tianmu Lake Institute of Advanced Energy Storage Technologies, Liyang, Jiangsu 213300, China*

[†] These authors contributed equally to this work.

A general feature of unconventional superconductors is the existence of a superconducting dome in the phase diagram as a function of carrier concentration[1]. For the simplest iron-based superconductor FeSe (with transition temperature $T_c \approx 8$ K)[2], its $T_c$ can be greatly enhanced by doping electrons via many routes[3-6], even up to 65 K in monolayer FeSe/SiTiO$_3$ (refs. 7-10). However, a clear phase diagram with carrier concentration for FeSe-derived

**superconductors is still lacking. Here, we report the observation of a series of discrete superconducting phases in FeSe thin flakes by continuously tuning carrier concentration through the intercalation of Li and Na ions with a solid ionic gating technique[11]. Such discrete superconducting phases are robust against the substitution of Se by 20% S, but are vulnerable to the substitution of Fe by 2% Cu, highlighting the importance of the iron site being intact. A complete superconducting phase diagram for FeSe-derivatives is given, which is distinct from other unconventional superconductors.**

The discovery of iron-based superconductors sparked a huge renaissance in exploring the mystery of unconventional superconductivity[12-15]. For the iron arsenides, there always exists a superconducting (SC) dome in the phase diagram, i.e., a continuous increase, maximum, and then decrease of the transition temperature $T_c$, with charge carrier doping, applied external pressure, or isovalent doping[14-16]. Such an SC dome is a common feature shared by all other unconventional superconductors, such as cuprate, heavy-fermion, and organic superconductors[1].

Iron selenide has become the star material recently since its $T_c$ can be significantly enhanced by electron doping and external pressure[3-6,17], even up to 65 K in monolayer FeSe/SrTiO$_3$, possibly with the assistance of phonons from the substrate[7-10]. A robust $T_c$ of ~32 K was first discovered in K$_x$Fe$_{2-y}$Se$_2$, followed by an occasional observation of trace SC at 44 K (ref. 3). However, attempts to ascertain a clear phase diagram have been hampered by the prevalence of phase separation which arises inevitably from the high-temperature synthesis routes[14,15]. An alternative strategy to obtain single-phase samples and avoid phase separation is to use a low temperature solvent such as liquid ammonia, but drawbacks include limited control over the intercalation of metals and sample instability[4,15].

The electrostatic gating technique provides a direct means of precisely controlling the doping content, and consequently the physical properties of a material[18]. In previous electrostatic gating of FeSe flakes, using ionic liquid as the gate dielectric, a jump in $T_c$ was reported at a certain gate voltage due to a Lifshitz transition, followed by a smooth increase of $T_c$ from ~30 K to 48 K (ref. 19). Two similar works successfully tuned insulating FeSe thin films into superconductors with $T_c$ increasing to about 40 K (refs. 20, 21). However, the charge accumulation induced by liquid ionic gating is confined to the immediate subsurface due to the Thomas-Fermi screening effect[22], and the doping level is limited due to sample damage at high gate voltages[19-21].

Here, by using the latest developed solid ionic gating technique[11], we continuously intercalate Li/Na into FeSe flakes and map out the entire phase diagram of metal-intercalated FeSe superconductors. Starting from the $T_c = 8$ K phase, two subsequent SC phases with $T_c$ of ~36 K and ~44 K are observed upon increasing the intercalation of Li, and finally the system enters an insulating state. Similar discrete SC phases are also found in Na$_x$FeSe. Considering the continuity of Li/Na intercalation, such discrete SC phases must be intrinsic and universal for FeSe derivatives obtained by metal intercalation. We find that these discrete SC phases is easily destroyed by disorder at the iron site in Li$_x$(Fe$_{0.98}$Cu$_{0.02}$)Se, but is insensitive to substitution at the selenium site. A comprehensive phase diagram of FeSe-derived superconductors is obtained in combination with previous surface spectroscopic studies.

Figure 1a shows the configuration of the fabricated solid ionic gating device. An FeSe thin flake with a thickness of 5~34 nm (9~61 monolayers) is sandwiched between the solid ionic substrate and an SiO$_2$ layer, as shown in Fig. 1b and 1c. We use 100 nm SiO$_2$ to insulate the electrodes from the substrate to avoid metal accumulation underneath the electrodes. This insulating layer is crucial to accurately determine the content of the intercalated metals. A window is left open on the SiO$_2$ layer to pattern the silver electrodes. When a back gate voltage, $V_G = 4$ V, is applied at low temperature (usually below 155 K), all the lithium or sodium ions in the substrate are still frozen in place. Upon slowly warming up the device above 165 K, these ions become mobile and are driven into the sample by the electric field. The ionic mobility can be precisely monitored by the leakage current, see Fig. S1 in the Supplementary Information.

We first examine the lightly doped region of Li$_x$FeSe. By carefully choosing the gating temperature and integrated leakage current, the lithium content $x$ can be finely tuned with a precision better than 1% (see Supplementary Information for more details). An FeSe flake with a thickness of 17.5 nm (31 monolayers, shown in Fig. 1d) is used, and the resistance as a function of temperature upon gating is plotted in Fig. 2a. Strikingly, an SC transition with $T_{c2} \approx 36$ K appears abruptly with negligible changes in the normal state resistance, coexisting with the SC transition at $T_{c1} = 8$ K (the bulk $T_c$ of FeSe). The transition at 8 K is then quickly suppressed and the normal state resistance continuously decreases upon increasing the doping level by the intercalation of Li. Meanwhile, the new SC transition remains at 36 K, but becomes sharper and sharper. Upon further increasing the Li content, another SC transition at $T_{c3}$ = 44 K sets in, coexisting with the 36 K phase. The 36 K SC phase is then suppressed until only the 44 K SC phase remains. As shown in Fig. 2b, the normal state resistance increases upon further gating, and the SC phase at 44 K gradually disappears until the sample eventually enters an insulating phase. Note that the observation of an insulating phase give a strong evidence of bulk intercalation, for the dopants must make it all the way across the sample to the leads, otherwise the surviving SC phase on that side of the sample would short out the leads. Since the gating process is very time consuming, the results shown in Fig. 2b are obtained on a 10 nm-thick sample.

To visualise these SC phases more clearly, we compute the derivative d$R$/d$T$ of each $R$-$T$ curve and plot the result as colour contours in Fig. 2c and 2d. The Li/Fe ratio dependence of $T_c$ shows step-like behaviour throughout the entire accessible doping range. The $T_{c2}$ phase emerges at a very low doping content ($x \approx 0.025$), and is replaced by the 44 K phase at $x \approx 0.17$. These results definitively demonstrate that the SC phases in Li$_x$FeSe are discrete rather than a continuous dome as observed in iron arsenides and other unconventional superconductors.

The previously reported 44 K phase was acquired by lithium intercalation with the assistance of organic molecules such as NH$_3$ (refs. 4, 15, 23). The Li content in Li$_{0.6(1)}$(ND$_2$)$_{0.2(1)}$(ND$_3$)$_{0.8(1)}$Fe$_2$Se$_2$ was determined by refining the neutron powder diffraction pattern[23]. To determine whether our $T_{c3}$ = 44 K phase is the same SC phase observed previously, we need to compare the Li content in them. During the gating process, the observation of two SC transitions suggests the coexistence of two SC phases. Once the SC transition at $T_{c1}$ = 8 K disappears, the SC transition at $T_{c2}$ = 36 K becomes sharper upon further gating – we assume that there exists only one single SC phase in the system when the transition at 36 K is sharpest. The same assumption is made for the $T_{c3}$ = 44 K phase. Based on this assumption, two repeated experiments on different samples give $x$ = 0.18 for the $T_{c2}$ = 36 K phase and $x$ = 0.35 for the $T_{c3}$ = 44 K phase by integrating the leakage current. The Li content in the SC phase with $T_{c3}$ = 44 K, $Li_{0.69(1)}Fe_2Se_2$, is consistent with that of $Li_{0.6(1)}(ND_2)_{0.2(1)}(ND_3)_{0.8(1)}Fe_2Se_2$ (ref. 23), which demonstrates that our determination of the Li content is reliable.

To establish whether the discrete SC phases are universal in other metal-intercalated FeSe, we carry out a similar experiment on a Na-based ionic substrate. As shown in Fig. 3, similar to $Li_xFeSe$, there are a series of discrete SC transitions in $Na_xFeSe$ at $T_{c1}$ = 8 K, $T_{c2}$ = 37 K, and $T_{c3}$ = 43 K. The observed 37 K and 43 K SC phases in $Na_xFeSe$ are consistent with the previously reported 36 K and 42 K SC phases in ammonia-free and ammonia-rich $Na_xFeSe(NH_3)_y$ (ref. 24). Our results strongly suggest that it is the Na content that determines the $T_c$ of $Na_xFeSe$. This is consistent with a recent theoretical calculation which shows that the $T_c$ is mainly determined by carrier concentration in intercalated FeSe, while different interlayer distances with the same doping level produce similar two-dimensional Fermi surfaces[25]. The observation of discrete superconducting phases in both $Li_xFeSe$ and $Na_xFeSe$ demonstrates that the absence of a continuous superconducting dome is intrinsic and universal in metal-intercalated FeSe.

To examine the robustness of this phase discreteness, we introduce disorder at the Se and Fe sites in FeSe. The substitution of S for Se has been proven to induce both disorder and chemical pressure, while the substitution of Cu for Fe introduces a negligible amount of electron doping, and most prominently, disorder[26]. The $T_c$s of FeSe$_{0.8}$S$_{0.2}$ and Fe$_{0.98}$Cu$_{0.02}$Se used here are 5 K and < 2 K, respectively (see Fig. S3 in the Supplementary Information), which agree well with the literature and confirm that Cu substitution has a much stronger effect on the superconductivity of FeSe. Figure 4a shows the resistance as a function of temperature during the gating process for FeSe$_{0.8}$S$_{0.2}$. The first SC transition at $T_{c1}$ becomes broader, and the onset $T_c$ gradually increases from 5 K to about 11 K, and then abruptly jumps to $T_{c2}' \approx 30$ K. No higher $T_c$ phase is observed upon further gating – the $T_{c3}$ phase is completely destroyed by the substitution of S for Se. These two discrete SC phases are clearly visible in the derivative in Fig. 4c for Li$_x$FeSe$_{0.8}$S$_{0.2}$. As shown in Fig. 4b, starting from a non-SC ($T_c$ < 2 K) Fe$_{0.98}$Cu$_{0.02}$Se thin flake, the superconductivity emerges upon gating and the onset $T_c$, shown in the inset, gradually increases to 23 K. The continuous evolution of $T_c$ in Li$_x$(Fe$_{0.98}$Cu$_{0.02}$)Se can be seen most clearly in Fig. 4d. This demonstrates that the substitution of only 2% Cu for Fe not only suppresses superconductivity, but also leads to a continuous change of $T_c$ in phase diagram, washing out the steps between the SC phases.

It is now possible to discuss the phase diagram of FeSe-derived superconductors in detail. The Fermi surface (FS) of bulk FeSe ($T_c$ = 8 K) consists of both electron and hole pockets; however, in FeSe-derived superconductors with a high $T_c$, such as $K_xFe_{2-y}Se_2$, $(Li_{0.8}Fe_{0.2})OHFeSe$, and monolayer FeSe/STO, only electron pockets at the Brillouin zone corner have been observed[27,28]. A clear evolution of FS topology can be seen in surface K-dosed FeSe films, where the hole pocket at the Brillouin center shrinks to zero upon increasing the potassium dosage, leaving the Fermi surface with only electron pockets at the M point[6,27,28]. This widely acknowledged Lifshitz transition of the FS topology is sketched in Fig. 5. In the recent electrostatic gating study on high-quality FeSe thin flakes, a steep $T_c$ jump was observed around the Lifshitz transition[19]. Above this transition, the increase of $T_c$ becomes more gradual and appears dome-shaped[19], consistent with the absence of a second Lifshitz transition and an enlarged electron Fermi surface at the M point upon further electron doping, as illustrated in Fig. 5.

Aside from surface doping, extensive studies focusing on the metal-intercalated FeSe superconductors that could produce a clear phase diagram with carrier density have not been previously performed. Based on our present study, the phase diagram of metal-intercalated FeSe, represented by $Li_xFeSe$, is plotted as the discrete red vertical lines in Fig. 5. This solves the puzzle of the observation of only certain $T_c$s for all the previously reported metal-intercalated FeSe superconductors, as well as a series of metal-intercalated superconductors also containing organic molecules, $A_xR_yFe_2Se_2$ (A = intercalation metals, R = organic molecule), even if intentionally varying the initial doping content[29]. Furthermore, our Fe- and Se-substitution experiments demonstrate that this intrinsic discreteness is easily destroyed by disorder on the Fe site. This is reminiscent of the situation in the cuprates, where a few percent disorder on the copper site has a huge effect on the physical properties including the superconductivity. The continuous evolution of $T_c$ in Li$_x$(Fe$_{0.98}$Cu$_{0.02}$)Se is also plotted in Fig. 5.

Our experiments constitute direct evidence of a discrete superconducting phase diagram as a function of carrier concentration in FeSe-derivatives, which is clearly distinct from all other observations on unconventional superconductors[1]. In the electrostatic gating process using ionic liquids, the large cations from the electrolyte accumulate at the surface of FeSe, only doping electrons into the subsurface of FeSe (ref. 19). Dosing potassium on the surface of FeSe thin film also only dopes electrons into the surface layer of FeSe (ref. 7-10). All these surface methods are quite different from the intercalation of Li and Na into the system driven by electric field which induces carriers into the system continuously. In the discrete SC phase diagram we report in Fig. 5, the transition between the different SC phases should be a first-order phase transition, which is very sensitive to disorder, possibly turning into a second-order transition. This may explains why the substitution of Cu for Fe not only suppresses superconductivity, but also leads to a continuous phase diagram. These results offer an insight into the mechanism of superconductivity in FeSe-derived superconductors.

**Methods**

FeSe single crystals were grown from a KCl-AlCl$_3$ flux. High-quality single crystals (typically 3×3×0.1 mm$^3$) were harvested after washing the product in deionized water. Resistance measurements of bulk samples show a sharp SC transition at 8 K. Sulphur- and copper-doped FeSe were synthesized following the same route, except that their raw materials (Fe$_{1.2}$Se$_{0.6}$Se$_{0.4}$ and (Fe$_{0.98}$Cu$_{0.02}$)$_{1.2}$Se) were heated to 1100 ℃ prior to single crystal growth for homogeneity. The actual compositions were determined to be FeSe$_{0.80(1)}$S$_{0.20(1)}$ and Fe$_{0.98(1)}$Cu$_{0.02(1)}$Se by inductively coupled plasma atomic emission spectroscopy (ICP-AES) and electron probe micro-analysis (EPMA).

Thin flakes were mechanically exfoliated by Scotch tapes and Polydimethylsiloxane (PDMS) embrace, and finally transferred to the solid ionic substrate under a microscope. The lithium-based substrate (Li$_{1+x+y}$Al$_x$(Ti,Ge)$_{2x}$Si$_y$P$_{3-y}$O$_{12}$) was purchased from OHARA INC, while the sodium-based substrate (Na$_{3.4}$Zr$_{1.8}$Mg$_{0.2}$Si$_2$PO$_{12}$) was synthesized following the procedure of ref. 30. The sodium-based substrate was finely polished before use to ensure a smooth surface with RMS roughness better than 10 nm within a 100×100 μm$^2$ area. Then, we used two kinds of shadow masks to make first a 100 nm SiO$_2$ protection layer and then the 200 nm-thick silver electrodes. The thickness of the samples was determined by Atomic Force Microscopy (AFM). All manipulations were performed in an inert-atmosphere glove-box or under vacuum to minimize oxidation.

Transport measurements were carried out in a Quantum Design Physical Properties Measurement System (PPMS) equipped with a Keithley 2400 source meter (with RMS noise about 20 pA, see Fig. S1) to apply the gate voltage, and a lock-in amplifier (SR830) to measure its resistance. A constant gate voltage ($V_G = 4$ V) was applied below 155 K first, then the device was slowly heated above 165 K for Li/Na intercalation and then cooled down, for one temperature cycle. The leakage current of the source meter shows that Li and Na ions in the substrates become mobile above 165 K. We set the upper temperature of the first tens of cycles to 170 K to observe the lightly doped region. After that, the upper temperature was gradually set to higher values to accelerate the intercalation process. It usually took 60 to 100 temperature cycles to complete the phase diagram.

**Acknowledgements:** We thank D. L. Feng, Y. Chen, J. P. Hu, N. Z. Wang, Y. Xu, and D. C. Peets for helpful discussions. This work was supported by the Ministry of Science and Technology of China (Grant No: 2015CB921401, 2016YFA0300503, and 2017YFA0303001), the Natural Science Foundation of China, Program for Professor of Special Appointment (Eastern Scholar) at Shanghai Institutions of Higher Learning, STCSM of China (Grant No. 15XD1500200), and China Postdoctoral Science Foundation (Grant No. 2016T90332).

**Author Contributions:** T.P.Y. and M.X.W. contributed equally to this work. Y.C.L., M.X.W. and T.P.Y. performed sample synthesis. Z.Z.Z. and W.Y. suggested the use of a sodium-based substrate and synthesized it. M.X.W. and T.P.Y. fabricated the nano-devices and performed the transport measurements with assistance from Z.Y.Z., X.Y.J., Q.L., B.L., Y.Y. and Y.Z. E.J.C. carried out the composition determination. S.Y.L. and T.P.Y. conceived the idea and designed the experiments. S.Y.L. and X.H.C. supervised the project. T.P.Y., S.Y.L., and X.H.C. analyzed the data and wrote the paper. All authors discussed the results and commented on the manuscript.

**Additional Information:** Correspondence and requests for materials should be addressed to X. H. Chen (chenxh@ustc.edu.cn) and S. Y. Li (shiyan_li@fudan.edu.cn).

**Competing financial interests:** The authors declare no competing financial interests.

**Figure 1 | Solid ionic gating of FeSe thin flakes. a**, Schematic structure of the solid ionic gating device. From the bottom to the top: silver back gate layer, finely polished Li/Na-ion substrate, pure or doped FeSe thin flake, 100 nm $SiO_2$ and four electrodes. The gate voltage $V_G$ is applied at low temperature (< 155 K) where all the ions in the substrate are frozen in place. A pico-ammeter is used to monitor the $V_G$ leakage current which is then used to calculate the metal content. The resistance was measured using a standard four-terminal configuration. **b**, Representative optical image of a fabricated device. Shadow masks are used here to eliminate any contamination from photoresist or developer. **c**, Atomic Force Microscope (AFM) topographic image of the dashed rectangular area shown in **b**. **d**, Cross-sectional profile of the thin flake along the red line in **c**, giving the thickness of the flake $t = 17.5$ nm, which corresponds to about 31 FeSe monolayers. The thickness of the samples used in the present study ranges from 5 to 30 nm.

**Figure 2 | Discrete superconducting phases of $Li_xFeSe$. a**, Resistance of the Li-intercalated FeSe flake (17.5 nm) as a function of temperature. This panel covers the underdoped regime with the normal state resistance monotonically decreasing upon gating. Colored dashed lines highlight transition temperatures where the $T_c$s (in this work, we use $T_c^{onset}$ as $T_c$) of several curves coincide. $T_{c1}$, $T_{c2}$, and $T_{c3}$ are superconducting (SC) phases at 8 K, 36 K, and 44 K, respectively. **b**, Resistance of a 10 nm FeSe flake in the overdoped regime with its normal state resistance gradually increasing upon gating. **c** and **d**, Colour contour plots of the derivatives of the *R-T*

curves in (**a**) and (**b**). We normalized the curves in (**b**) at 80 K before differentiating to see the SC transitions more clearly.

**Figure 3 | Discrete superconducting phases of Na$_x$FeSe. a** and **b**, similar to those of Fig. 2, are the resistance of Na-intercalated FeSe flakes as a function of temperature. The sample thicknesses are 18 nm for (**a**) and 8.5 nm for (**b**), respectively. Vertical broken lines highlight transition temperatures where the $T_c$s of different curves coincide. $T_{c2}$ is observed throughout the entire doping region before the sample goes into the insulating phase. **c** and **d**, Colour contour plots of the derivatives of the *R-T* curves in (**a**) and normalized (**b**).

**Figure 4 | The effect of disorder at Se and Fe sites. a** and **b**, The gating dependence of *R-T* curves for FeSe$_{0.8}$S$_{0.2}$ (34 nm thick) and Fe$_{0.98}$Cu$_{0.02}$Se (21.5 nm thick). **c** and **d** are their respective dR/dT colour contour plots. We note that the $T_{c2}'$ of Li$_x$Fe(Se$_{0.8}$S$_{0.2}$) may not necessarily correspond to $T_{c2}$ in Li$_x$FeSe. The inset in (**b**) expands the SC onset region. Red arrows denote the $T_c$ for each curve and are included in (**d**) as open red circles.

**Figure 5 | Phase diagram of FeSe-derived superconductors.** The vertical red lines denote the discrete SC phases of metal-intercalated FeSe superconductors, as represented by Li$_x$FeSe. The different Fermi surface topologies of pristine and doped FeSe are superimposed. The blue curve and diamonds represent the emergence of a continuous SC dome with 2% substitution of Fe by Cu. The yellow area on the right represents the insulating phase in the heavily overdoped region.

**Figure 1**

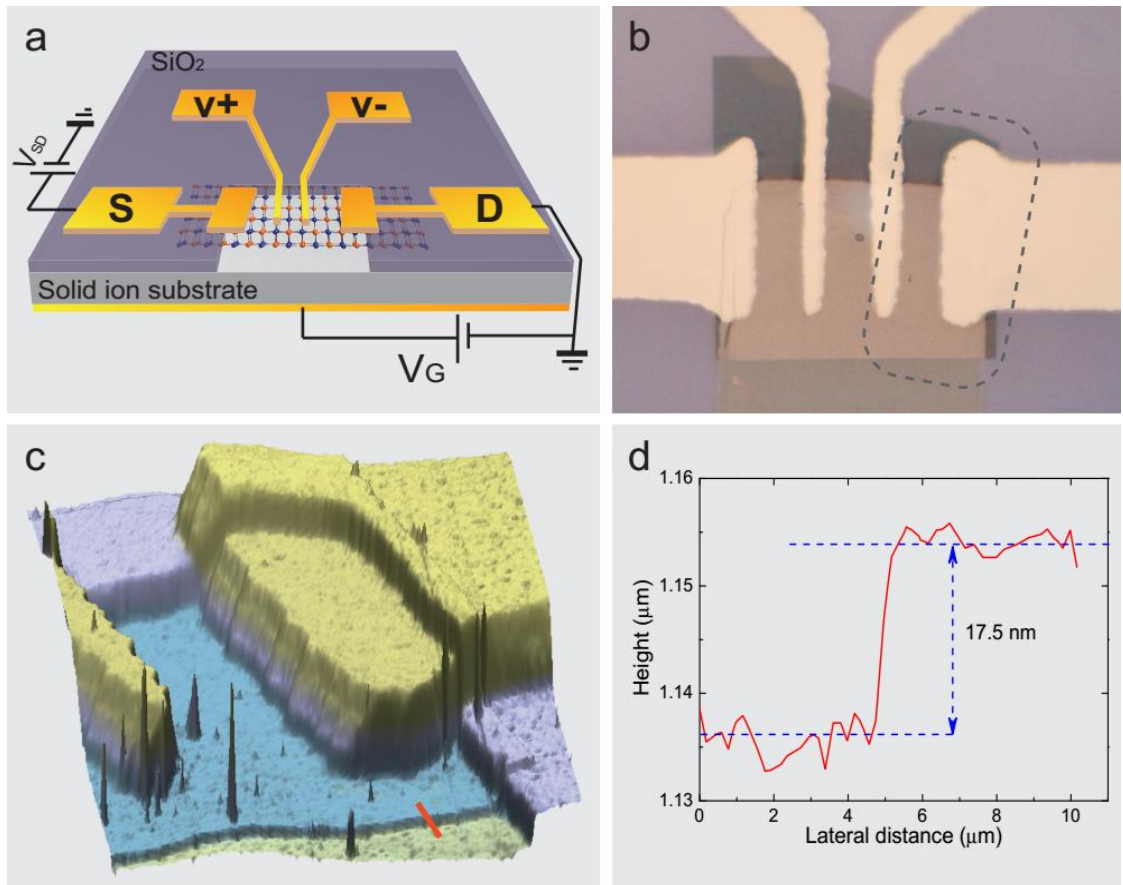

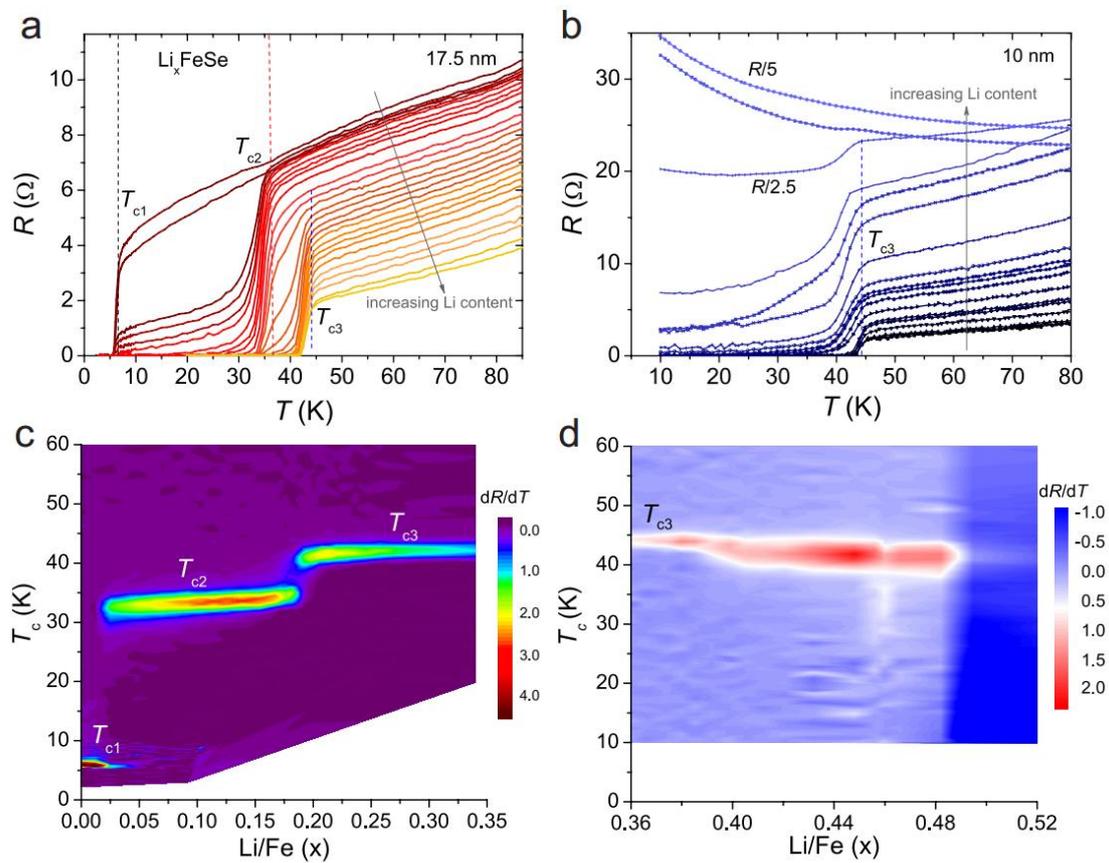

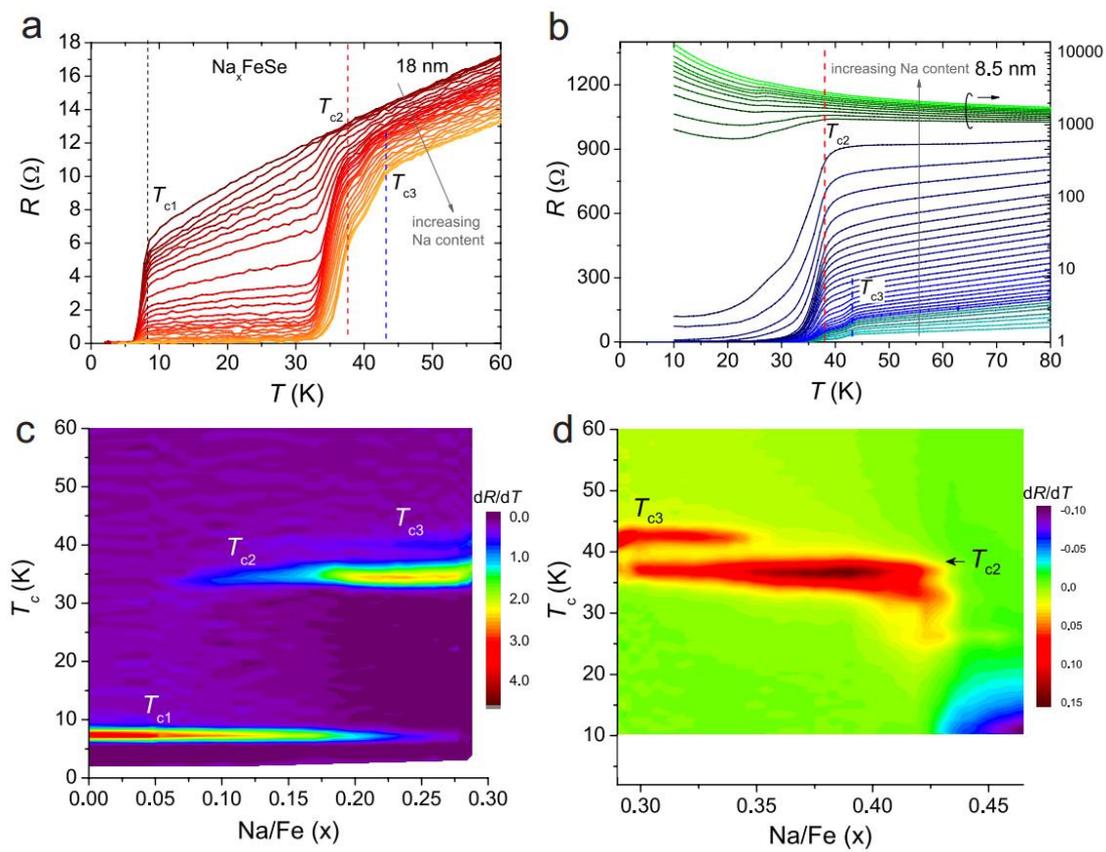

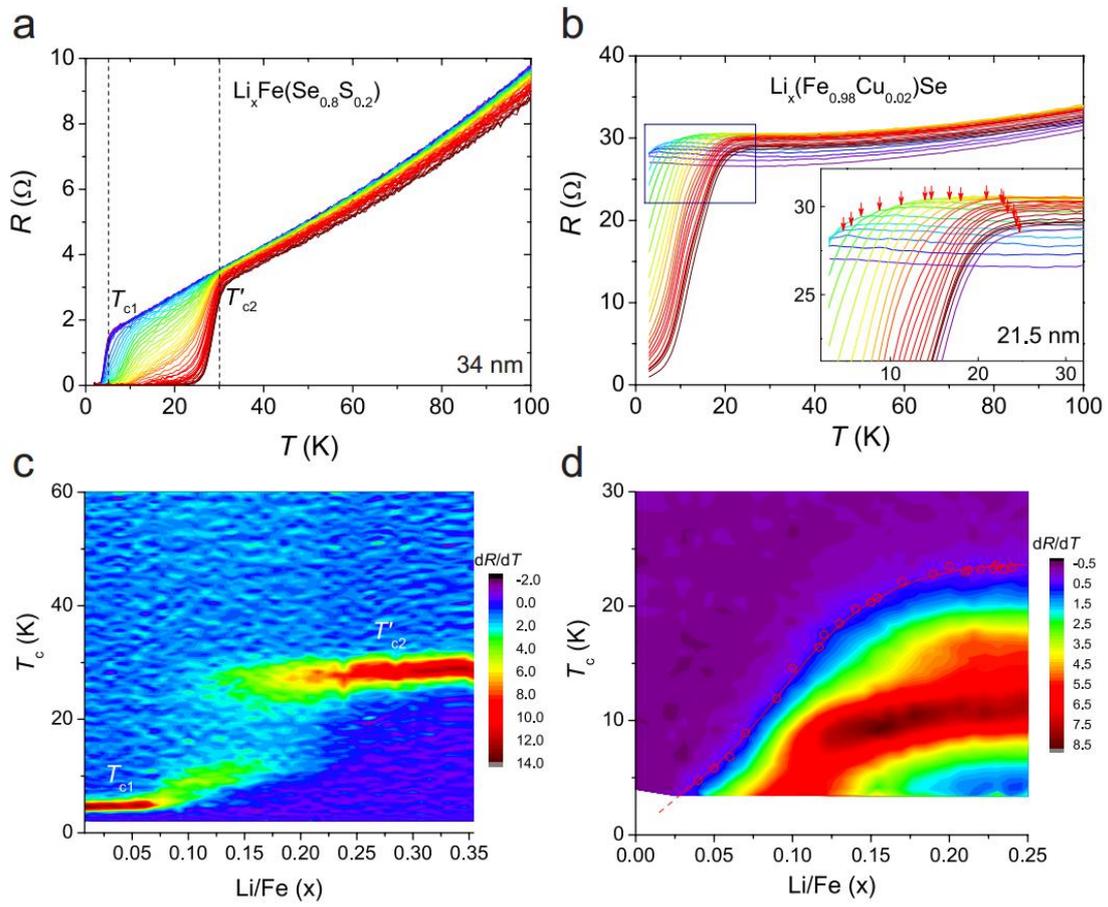

**Figure 5**

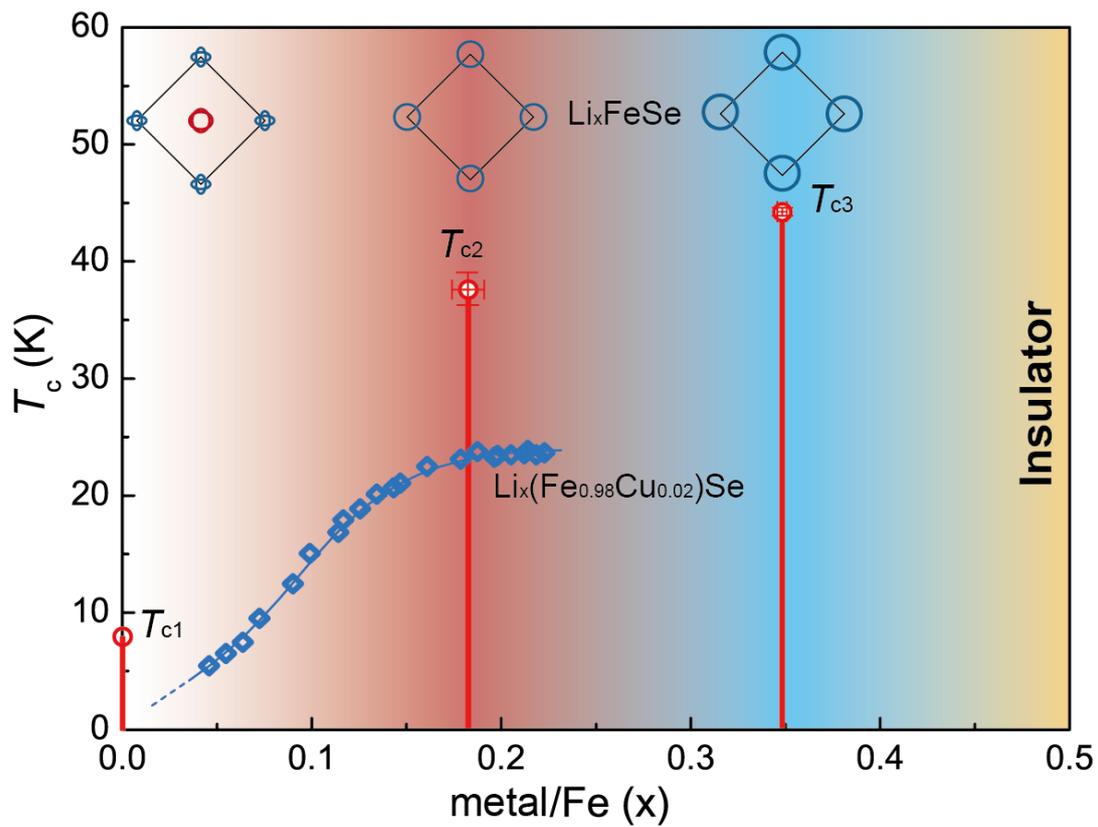

# Supplementary Information for "Discrete superconducting phases in FeSe-derived superconductors"

T. P. Ying[†], M. X. Wang[†], Z. Y. Zhao, Z. Z. Zhang, X. Y. Jia, Y. C. Li, Q, Li, B. Lei, Y. Yu, E. J. Cheng, Z. H. An, Y. Zhang, W. Yang, X. H. Chen[*], S. Y. Li[*]

[*]X. H. Chen (chenxh@ustc.edu.cn) and S. Y. Li (shiyan_li@fudan.edu.cn)

**Contents**

  **1. Determination of the leakage current**

  **2. Lithium content determination**

  **3. The resistance of bulk FeSe, FeSe$_{0.8}$S$_{0.2}$, and Fe$_{0.98}$Cu$_{0.02}$Se single crystals**

1. **Determination of the leakage current**

It is vital to insulate the electrodes from the substrate if we want to get an accurate Li/Fe ratio. A piece of Li-based substrate with no sample on it is used as a control experiment. As shown in Fig. S1, the leakage current is negligible for the red curve below 225 K, where no FeSe is used. It means that the 100 nm $SiO_2$ is sufficient to insulate the electrodes from the substrate. So the highest temperatures in the intercalation procedure are kept below 225 K to get an accurate Li content. With the small noise in the leakage current (about 20 pA), we can observe the onset of the intercalation for $Li_xFeSe$ at 165 K (blue curve).

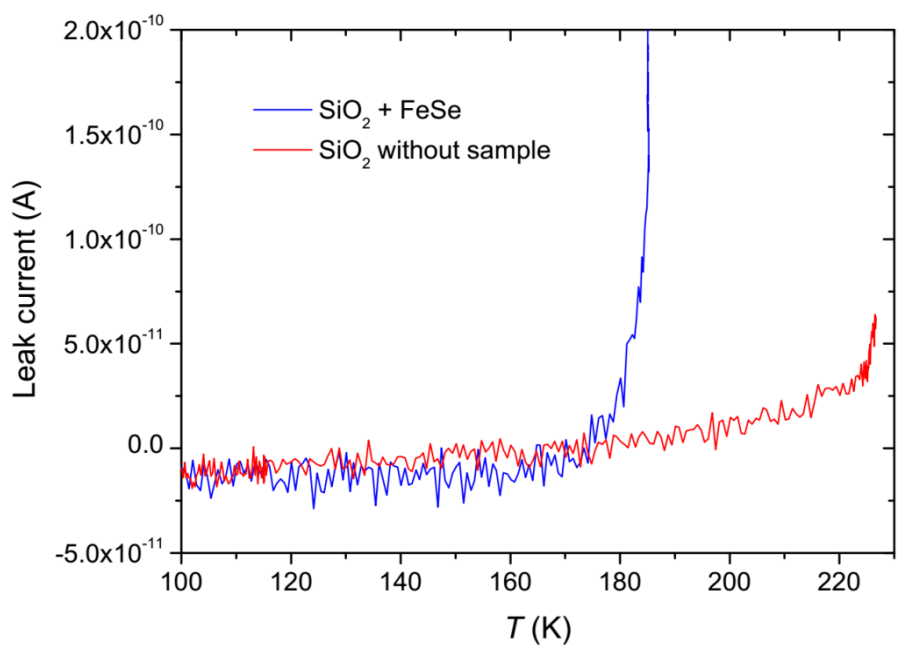

**Figure S1 | The temperature dependence of leakage current of devices with and without a sample.** As shown in the blue curve, the leakage current remains constant below 165 K, which means that all the Li ions in the substrate are frozen in place. Above 165 K, the current increases rapidly. The red curve is a control experiment where only a 100 nm $SiO_2$ evaporated blank substrate is used. The leakage current is

negligible below 225 K. We attribute the current observed in the blue curve to the intercalation of lithium ions into FeSe, and the 100 nm $SiO_2$ can effectively insulate the electrode from the substrate for temperatures below 225 K.

## 2. Lithium content determination

Figure S2 shows the time dependence of the leakage current and Li/Fe ratio for one conventional $Li_xFeSe$ device. Each blue peak is formed by warming the device above 165 K and then cooling it. The maximum of each peak depends on the end temperature of the warming process (from 170 K to 225 K). The mobility of Li ions in the substrate is determined by the temperature. The higher the temperature, the larger the peak value will be. Due to depletion of the lithium ions in the substrate, the end temperature should be set to a higher value after several cycles. It usually takes 60 to 100 cycles to complete the doping of a typical device, allowing us to fine-tune the lithium content with a precision of 1%.

We do not specify the sodium content in $Na_xFeSe$, not only because of the roughness of the Na-based substrate, where 100 nm $SiO_2$ cannot completely insulate the electrodes, but also the inhomogeneous distribution of the Na ions within the FeSe thin flake. The horizontal axes of Fig. 3c and 3d are referenced from that of Fig. 2 by assuming that their highest-$T_c$ phases have a similar doping level of alkali metal.

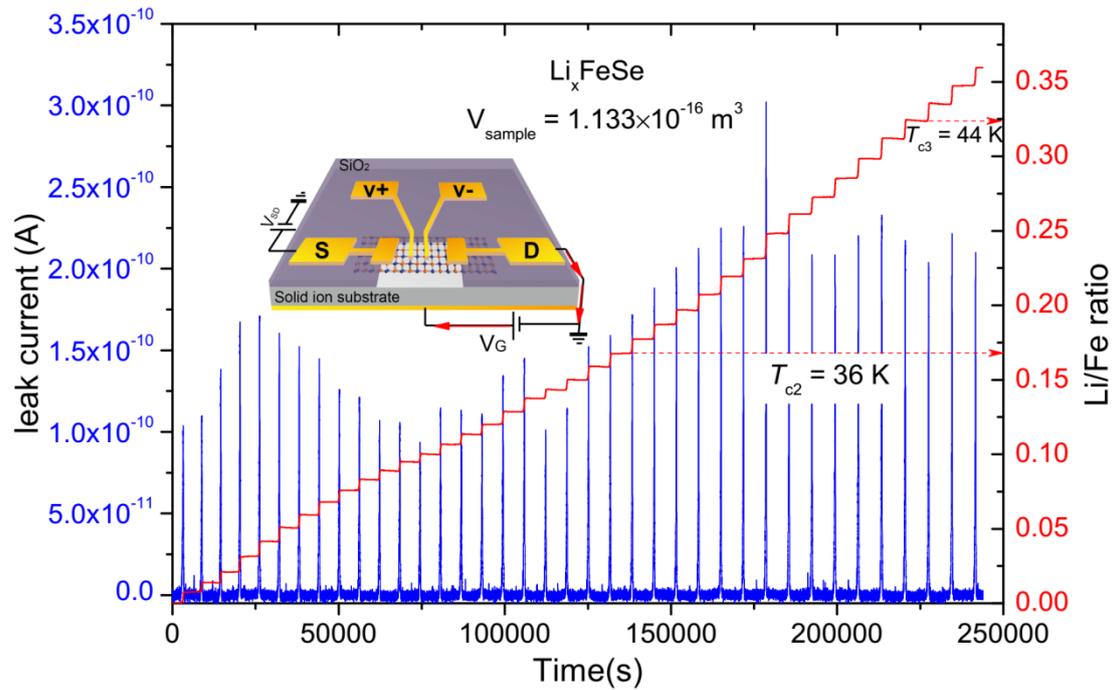

**Figure S2 | The leakage current and its integral.** The gating leakage current is schematically shown as the solid red arrows in the inset. Blue peaks are the time dependence of the leakage current and the red stair-like line is its integral. Two red dashed arrows correspond to the Li/Fe ratios whose corresponding curves in Fig. 2 possess the sharpest transitions for the $T_{c2}$ = 37 K and $T_{c3}$ = 44 K superconducting phases.

## 3. The resistance of bulk FeSe, FeSe$_{0.8}$S$_{0.2}$, and Fe$_{0.98}$Cu$_{0.02}$Se single crystals

*R-T* curves of bulk FeSe, FeSe$_{0.8}$S$_{0.2}$, and Fe$_{0.98}$Cu$_{0.02}$Se single crystals are shown in Fig. S3. Their resistance is normalized at 285 K for comparison. A sharp $T_c$ at 8 K can be seen for FeSe. The 20% S-doped sample exhibits a $T_c$ at 5 K, while 2% Cu doping at the Fe site effectively suppresses the $T_c$ to below 2 K.

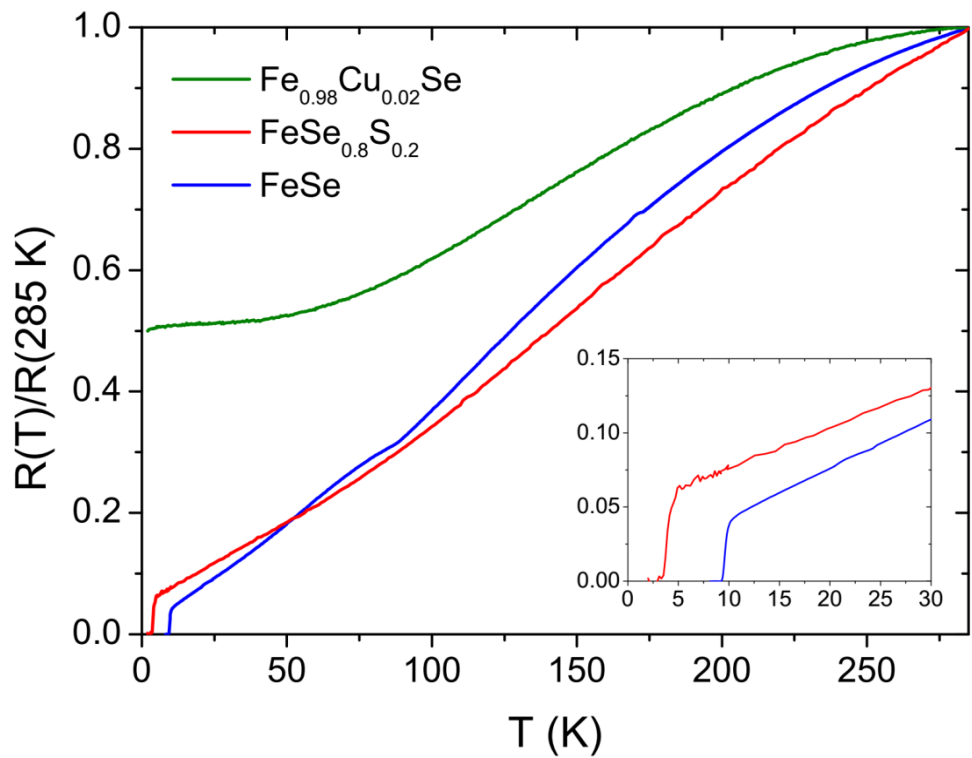

**Figure S3 | The temperature dependence of resistance for bulk FeSe, FeSe$_{0.8}$S$_{0.2}$ and Fe$_{0.98}$Cu$_{0.02}$Se.** The substitution of S at the Se site leads a slight decline of $T_c$ to 5 K in FeSe$_{0.8}$S$_{0.2}$, while the doping of 2% Cu at Fe site effectively suppresses the $T_c$ to below 2 K. Inset is an enlargement of the SC transition.